\documentclass[prd,preprint,showpacs,showkeys,amsmath,amssymb,eufrak]{revtex4}

\usepackage{psfrag}
\usepackage{graphicx}

\newcommand{\vc}[1]{{\boldsymbol{#1}}}  

\begin{document}

\title{On Gravitational Radiation in Quadratic $f(R)$ Gravity}

\author{Joachim N\"af} 
\email{naef@physik.uzh.ch}

\author{Philippe Jetzer}

\affiliation{Institut f\"{u}r Theoretische Physik, Universit\"{a}t Z\"{u}rich,
Winterthurerstrasse 190, CH-8057 Z\"{u}rich, Switzerland.}

\begin{abstract}

We investigate the gravitational radiation 
emitted by an isolated system for gravity 
theories with Lagrange density $f(R) = R + aR^2$. As a formal result we obtain 
leading order corrections to the quadrupole formula in General Relativity. 
We make use of the analogy of $f(R)$ theories with scalar--tensor theories, 
which in contrast to General Relativity feature an additional scalar degree of freedom. 
Unlike General Relativity, where the leading order gravitational radiation is produced by 
quadrupole moments, the additional degree of freedom predicts gravitational 
radiation of all multipoles, in particular monopoles and dipoles, 
as this is the case for the most alternative 
gravity theories known today. An application to a hypothetical binary pulsar
moving in a circular orbit yields the rough limit $a \lesssim 1.7\cdot10^{17}\,\mathrm{m}^2$ 
by constraining the dipole power to account at most for $1\%$ of the quadrupole power 
as predicted by General Relativity.

\end{abstract}


\maketitle


\section{Introduction}

One of the most impressive endorsements of General Relativity Theory (GR) is the 
agreement of the predictions of the famous quadrupole formula for gravitational 
radiation with indirect measurements of the energy loss of binary pulsars. It 
is thus natural to test modified gravity theories by deriving the 
corrections to the quadrupole formula and comparing them with experimental data. 
For many types of theories this has been done in the past \cite{WILL}. Though 
this problem is still open for metric $f(R)$ theories with an action
\begin{equation}\label{action}
S = \frac{c^3}{16\pi G}\int f(R)\sqrt{-g}\,d^4x + S_M,
\end{equation}
where in contrast to GR the Einstein--Hilbert Lagrangian density is replaced by 
a nonlinear function $f(R)$. We will always assume the function $f$ to be smooth. 
$S_M$ is the standard matter action. 
In the past years, this type of theories has become very popular to 
heuristically gain insight in the problem of dark energy. 
For an overview one may consult e. g. \cite{caplau,STR2} and references therein. 
It is also worth mentioning that quadratic corrections $aR^2$ 
to the Einstein--Hilbert action have been considered in the context of 
inflationary cosmology already in the 1980's 
\cite{staro,vilen}.

In this article we prepare the way to investigate the energy emission of 
binary systems by gravitational radiation. The basic equations of $f(R)$ 
gravity are given in Section II. For our purposes 
it will be convenient to work in the scalar tensor formulation of quadratic 
$f(R)$ gravity. In Section III we employ the linearised field equations of 
quadratic $f(R)$ gravity to derive the weak gravitational fields emitted by 
a localised source and expand them into multipoles. The linearised $f(R)$ gravity 
has been investigated for example in \cite{caplau,CORDA1,CORDA2,CORDA3,begair}. For a treatment 
of linearised scalar tensor theories see \cite{WILL,wagon}. In Section IV we dwell 
on the energy--momentum complex in quadratic $f(R)$ gravity as an analogue to the 
Landau--Lifshitz complex in GR. The leading order correction to the 
quadrupole formula in terms of momenta of the energy--momentum 
tensor is derived in in Section V. In Section VI we finally illustrate the correction 
with an application to binary systems in circular orbits. 

Notational conventions: Greek letters denote space time indices and 
range from $0$ to $3$, whereas Latin letters denote space indices and range 
from $1$ to $3$. We take the sum over repeated indices within a term.


\section{The Field Equations}

Consider a 4-dimensional pseudo Riemannian manifold with metric 
$g_{\mu\nu}$ of signature $(-,+,+,+)$. We write $g = \det{g_{\mu\nu}}$ 
and denote the Ricci tensor of $g_{\mu\nu}$ by $R_{\mu\nu}$. 
The variation of the action (\ref{action}) with respect to the metric 
yields the Euler--Lagrange equations
\begin{equation}\label{eullag}
f'(R)R_{\mu\nu} - \frac{1}{2}f(R)g_{\mu\nu} - 
\nabla_{\mu}\nabla_{\nu}f'(R) + g_{\mu\nu}\square_g f'(R) = 
\frac{8\pi G}{c^4}T_{\mu\nu},
\end{equation}
where $R = g^{\mu\nu}R_{\mu\nu}$, 
$T_{\mu\nu} = (-2c/\sqrt{-g})(\delta S_M/\delta g^{\mu\nu})$ is the energy-momentum 
tensor, $c$ the vacuum speed of light, $G$ Newton's constant, 
$\nabla_{\mu}$ the covariant derivative for $g_{\mu\nu}$ and 
$\square_g = \nabla^{\mu}\nabla_{\mu}$. Taking the trace of (\ref{eullag}) we obtain
\begin{equation}\label{eullagtr}
3\square_g f'(R) + f'(R)R - 2f(R) = \frac{8\pi G}{c^4}T,
\end{equation}
where $T$ is the trace of $T_{\mu\nu}$.
We now assume 
\begin{equation}\label{f(R)}
f(R) = R + aR^2
\end{equation}
and make use of the equivalence between $f(R)$ gravity and scalar tensor 
theory by defining the scalar field $\phi := f'(R)$. This identification 
is feasible since $f''(R) \neq 0$ holds for our choice of 
$f(R)$, and $f'(R)$ is thus invertible. We define the scalar field 
$\varphi$ by $\phi = 1 + 2a\varphi$, where we have chosen the asymptotic value 
such that a renormalisation of Newton's constant is redundant. Then the 
equations (\ref{eullag}) and (\ref{eullagtr}) are equivalent to
\begin{eqnarray}\label{eullag1}
R_{\mu\nu} - \frac{1}{2}R g_{\mu\nu} & = &
\frac{1}{1+2a\varphi}\left[\frac{8\pi G}{c^4}T_{\mu\nu} + 
a\left(2\nabla_{\mu}\nabla_{\mu}\varphi - 2g_{\mu\nu}\square_g \varphi
 - \frac{1}{2}g_{\mu\nu}\varphi^2\right)\right]
\\ \label{eullag2}
\square_g \varphi& = & \frac{4\pi G}{3ac^4}T + \frac{1}{6a}\varphi.
\end{eqnarray}
The field $\varphi$ thus has the effective mass $\hbar/(c\sqrt{6a})$. 
Since we aim at the investigation of isolated systems at the scale 
of the Solar System, we expect the theory derived from (\ref{f(R)}) to be a small 
perturbation of General Relativity. Thus the dimensionless quantity $aR$ should be 
small compared to $1$. This still holds if the parameter $a$ varies appropriately 
with the scalar curvature and therefore the local matter density.
For our quadratic model this fact reflects the concept 
of the chameleon effect \cite{KhWe}, which states the possibility that the Compton 
wave length $\lambda = \sqrt{6a}$ of the field $\varphi$ is smaller or larger 
in regions with higher or lower matter density, respectively. In our specific 
case however, we assume the Compton wave length to be constant. In this sense 
theory should be viewed as a local effective field theory which is valid for a 
certain range of the parameters.


\section{Gravitational Radiation in $f(R)$ gravity}

Consider weak perturbations of the Minkowski spacetime metric
$\eta_{\mu\nu} = \mathrm{diag}(-1,1,1,1)$. The metric can be written as
\begin{equation}\label{}
g_{\mu\nu} = \eta_{\mu\nu} + h_{\mu\nu},
\end{equation}
where the coefficients of the perturbation satisfy $|h_{\mu\nu}| \ll 1$. In what 
follows the indices are raised and lowered by $\eta_{\mu\nu}$. For the 
field $\phi$ we have already chosen the asymptotic value $1$, of which 
$2a\varphi$ is the perturbation. Moreover, the field equation (\ref{eullag2}) is 
inhomogeneous linear in $\varphi$, so that the linearisation in the 
perturbations is simply achieved through the replacement of 
$\square_g$ by $\square_\eta$. Let $h = h^{\mu}_{\mu}$, define
\begin{equation}
\gamma_{\mu\nu} = h_{\mu\nu} - \frac{1}{2}h\eta_{\mu\nu} -2a\varphi\eta_{\mu\nu} 
\end{equation}
and choose the gauge
\begin{equation}\label{gauge}
\gamma_{\mu\nu}^{\phantom{\mu\nu},\nu} = 0.
\end{equation}
Up to linear order in $h_{\mu\nu}$ and $\varphi$, equation (\ref{eullag1}) can 
then be written as
\begin{equation}\label{lingmunu}
\square_\eta\gamma_{\mu\nu} = \frac{16\pi G}{c^4}T_{\mu\nu}.
\end{equation}
In the slow motion approximation and at large distances from the localised 
sources, a special solution of (\ref{lingmunu}) can be 
derived in analogy with the GR case. Since $T_{\mu\nu}$ is divergence free, 
we can to express the spatial components of $\gamma_{\mu\nu}$ in terms of 
the quadrupole momenta of $T^{00}$. Thus we obtain
\begin{equation}\label{lingmunusol}
\gamma^{ij}(t,\vc{x}) = 
\frac{2G}{c^6}\frac{1}{|\vc{x}|}\frac{\partial^2}{\partial t^2}
\int_{\mathbb{R}^3} d^3x'T^{00}(t-|\vc{x}|/c,\vc{x'})x'^ix'^j.
\end{equation}
For the field $\varphi$ we write equation (\ref{eullag2}) as
\begin{equation}\label{eullag2l}
\square_\eta \varphi - \alpha^2\varphi = \frac{8\pi G\alpha^2}{c^4}S,
\end{equation}
where $\alpha := 1/\sqrt{6a}$ and 
\begin{eqnarray}
S = T\left[1+\frac{1}{c^2}\left(3W+\frac{2}{3\alpha^2}V\right)\right]
+\frac{1}{8\pi G}\left[\frac{1}{3\alpha^4}(\nabla V)^2 + UV + VW\right]
\end{eqnarray}
is the source $T$ (to leading order) extended by the terms which are quadratic in the perturbations. These are 
expressed in terms of the Newtonian and post Newtonian potentials $U$, $W$ and $V$, which
are given by \cite{phich,clifton}
\begin{eqnarray}
U(\vc{x},t) & = & \frac{4G}{3c^2}\int\frac{{}^{(-2)}
T^{00}(\vc{x}',t)}{\left|\vc{x}-\vc{x}'\right|}d^3x', \\
W(\vc{x},t) & = & \frac{2G}{3c^2}
\int\frac{{}^{(-2)}T^{00}(\vc{x}',t)}{\left|\vc{x}-\vc{x}'\right|}\left(1-e^{-\alpha\left|\vc{x}-\vc{x}'\right|}\right)d^3x', \nonumber \\
V(\vc{x},t) & = & \frac{2G\alpha^2}{c^2}
\int\frac{{}^{(-2)}T^{00}(\vc{x}',t)
e^{-\alpha\left|\vc{x}-\vc{x}'\right|}}{\left|\vc{x}-\vc{x}'\right|}d^3x'. \nonumber
\end{eqnarray}
Here ${}^{(-2)}T^{00}$ is the leading order time--time component of $T^{\mu\nu}$. 

A special solution of (\ref{eullag2l}) is the convolution of the source 
with the Green's function of the Klein--Gordon equation \cite{morfesh},
\begin{equation}
\mathcal{G}(t,\vc{x}) = -\frac{1}{4\pi}\left[\frac{\delta\left(t-
|\vc{x}|/c\right)}{|\vc{x}|} - 
\frac{\alpha J_1\left(\alpha c \sqrt{t^2 - (|\vc{x}|/c)^2}\right)
\theta\left(t-
|\vc{x}|/c\right)}{\sqrt{t^2 - (|\vc{x}|/c)^2}}
\right],
\end{equation}
where $J_1$ the Bessel function of first order and the Dirac, and Heaviside, distribution kernel is denoted by $\delta$, and $\theta$, respectively. Then
\begin{eqnarray}\label{scasol}
\varphi(t,\vc{x}) & = & -\frac{2G\alpha^2}{c^4}\Bigg[
\int_{\mathbb{R}^3}d^3x'\frac{S\left(t-
|\vc{x}-\vc{x'}|/c,\vc{x'}\right)}{|\vc{x}-\vc{x'}|} \\
& & \quad {}- \int_{-\infty}^{t-|\vc{x}-\vc{x'}|/c}dt'
\int_{\mathbb{R}^3}d^3x'\frac{\alpha 
J_1\left(\alpha c \sqrt{(t-t')^2 - (|\vc{x}-\vc{x'}|/c)^2}\right)}
{\sqrt{(t-t')^2 - (|\vc{x}-\vc{x'}|/c)^2}}S(t,\vc{x'})
\Bigg]. \nonumber
\end{eqnarray}
If the source emits a single pulse, the field $\varphi$ observed at a distance 
$|\vc{x}|$ consists of this pulse diminished by the factor $1/|\vc{x}|$ given 
by the first term on the right hand side of (\ref{linphisol}), and a wake represented 
by the second term. The scalar mode is thus dispersive; for a discussion of the dispersion 
of plane scalar waves see \cite{CORDA1,begair,wagon}.

After the substitution 
\begin{equation}
t'=t-\frac{|\vc{x}-\vc{x'}|}{c}\sqrt{1+\frac{s^2}{\alpha^2|\vc{x}-\vc{x'}|^2}}, 
\quad dt'=-\frac{1}{\alpha^2c|\vc{x}-\vc{x'}|}\frac{s}{\sqrt{1+\frac{s^2}{\alpha^2|\vc{x}-\vc{x'}|^2}}}\,ds
\end{equation}
in the second term of the right hand side of (\ref{scasol}), the solution can be written as
\begin{eqnarray}\label{linphisol}
\varphi(t,\vc{x}) & = & \frac{G}{3ac^4}\int_{\mathbb{R}}ds\left[J_1(s)\theta(s) - \delta(s)\right]
\int_{\mathbb{R}^3}d^3x'
\frac{S\left(t-|\vc{x}-\vc{x'}|\sqrt{1+\frac{6as^2}{|\vc{x}-\vc{x'}|^2}}/c,\vc{x'}\right)}{|\vc{x}-\vc{x'}|\sqrt{1+\frac{6as^2}{|\vc{x}-\vc{x'}|^2}}}.
\end{eqnarray}
We assume the source $S$ to be a smooth function of time. In the limit $a \to 0$, the 
spatial integral in (\ref{linphisol}) is independent of $s$, and because of 
$\int_0^\infty J_1(s)ds = 1$ we have $\lim_{a\to0}a\varphi = 0$.

Let $r := |\vc{x}|$ and $\vc{n} := \vc{x}/r$. At large distances $r$ from isolated and 
slowly moving sources, we can expand (\ref{linphisol}) into multipoles. Since the scalar 
contribution to the energy flux (equation (\ref{flux}) below) is quadratic in $\varphi$ 
and $\gamma_{\mu\nu}$, and moreover, the leading order radiation predicted by GR is 
produced by quadrupole moments, we need an expansion for $\varphi$ taking into account 
up to hexadecapole moments. We therefore derive the fourth order Taylor polynom of the 
integrand in (\ref{linphisol}) around the origin in the variable $\vc{x}'$, while ignoring 
the explicit dependence of $S$ on $\vc{x}'$. Consider a source with an extension 
characterised by a typical length $d \ll r$ and moving slowly at a velocity characterised 
by a typical frequency $\omega \ll c/d$. The polynomial can be written as a sum of terms of 
approximate order $\mathcal{O}\left(\frac{d^n}{r^mc^l}\frac{\partial^l}{\partial t^l}S\right) 
= \mathcal{O}\left(\frac{d^n\omega^l}{r^mc^l}\right)$, $m+l = n$. In fact the expansion 
coefficients of the monomials $x'^{i_1}\cdots x'^{i_n}$ are not proportional to $r^{-m}$, but 
depend on $r$ also by means of the function
\begin{equation}
p(s) := \left(1+\frac{6as^2}{r^2}\right)^{-1/2}.
\end{equation}
For convenience, we will ignore the function $p$ when making use of the Landau Symbol $\mathcal{O}$. 
We define the retarded time 
\begin{equation}
\tau(t,s) := t-\frac{r}{p(s)c}
\end{equation}
and the distribution kernel
\begin{equation}
q(s) := p(s)[J_1(s)\theta(s) - \delta(s)].
\end{equation}
We are then left with
\begin{eqnarray}\label{linphisolexp}
\varphi(t,\vc{x}) & = & \frac{G}{3ac^4r}\int_\mathbb{R}ds\,q(s)\int_{\mathbb{R}^3} d^3x'\,
\Big[1 + F_i(s)x'^i + F_{ij}(s)x'^ix'^j\\ 
& & \quad {}+ F_{ijk}(s)x'^ix'^jx'^k + F_{ijkl}(s)x'^ix'^jx'^kx'^l\Big]S\left(\tau(t,s),\vc{x}'\right) + 
\mathcal{O}\left(\frac{d^5\omega^m}{r^nc^m}\right),\nonumber
\end{eqnarray}
($n+m = 5$), where
\begin{eqnarray}\label{func}
F_i(s) &:=& n_i\left[\frac{p^2(s)}{r} + \frac{p(s)}{c}\frac{\partial}{\partial t}\right],\\
F_{ij}(s) &:=& n_in_j\left[\frac{3p^4(s)}{2r^2} + \frac{p^3(s)}{rc}\frac{\partial}{\partial t} + \frac{p^2(s)}{2c^2}\frac{\partial^2}{\partial t^2}\right] - \delta_{ij}\left[\frac{p^2(s)}{2r^2} + \frac{p(s)}{2rc}\frac{\partial}{\partial t}\right],\nonumber\\
F_{ijk}(s) &:=& n_in_jn_k\left[\frac{5p^6(s)}{2r^3} + \frac{p^5(s)}{2r^2c}\frac{\partial}{\partial t} + \frac{p^4(s)}{2rc^2}\frac{\partial^2}{\partial t^2} + \frac{p^3(s)}{6c^3}\frac{\partial^3}{\partial t^3}\right] \nonumber\\
& & \quad {}- n_i\delta_{jk}\left[\frac{3p^4(s)}{2r^3} + \frac{p^3(s)}{2r^2c}\frac{\partial}{\partial t} + \frac{p^2(s)}{rc^2}\frac{\partial^2}{\partial t^2}\right],\nonumber\\
F_{ijkl}(s) &:=& n_in_jn_kn_l\left[\frac{35p^8(s)}{8r^4} + \frac{35p^7(s)}{8r^3c}\frac{\partial}{\partial t} + \frac{15p^6(s)}{8r^2c^2}\frac{\partial^2}{\partial t^2} - \frac{p^5(s)}{12rc^3}\frac{\partial^3}{\partial t^3} + \frac{p^4(s)}{24c^4}\frac{\partial^4}{\partial t^4}\right] \nonumber\\
& & \quad {}- n_in_j\delta_{kl}\left[\frac{15p^6(s)}{4r^4} + \frac{15p^5(s)}{4r^3c}\frac{\partial}{\partial t} + \frac{3p^4(s)}{2r^2c^2}\frac{\partial^2}{\partial t^2} - \frac{p^3(s)}{12rc^3}\frac{\partial^3}{\partial t^3}\right]\nonumber\\
& & \quad {}- \delta_{ij}\delta_{kl}\left[\frac{3p^4(s)}{8r^4} + \frac{3p^3(s)}{8r^3c}\frac{\partial}{\partial t} + \frac{p^2(s)}{8r^2c^2}\frac{\partial^2}{\partial t^2}\right].\nonumber
\end{eqnarray}
In what follows we drop the post Newtonian source terms that are quadratic in the perturbation fields. 
Note that these contributions would lead to corrections in the energy emission formula which are quadratic in $G$. 
Therefore we will neglect them and proceed with $T^{\mu\nu}$ as the main contribution 
to the source, i.e. $S = \eta_{\mu\nu}T^{\mu\nu}$. If we consider a perfect, non viscous fluid with 
mass density $\rho$, pressure $\mathcal{P}$ and velocity field $\vc{v} = (v^1,v^2,v^3)$, 
we have
\begin{eqnarray}\label{}
T^{00}\left(t,\vc{x}\right) & = & c^2\left[\rho\left(t,\vc{x}\right) + \mathcal{O}\left(c^{-2}\right)\right], \\ 
T^{0i}\left(t,\vc{x}\right) & = & c\left[\rho\left(t,\vc{x}\right) v^i\left(t,\vc{x}\right) + \mathcal{O}\left(c^{-2}\right)\right], \nonumber \\ 
T^{ij}\left(t,\vc{x}\right) & = & \rho\left(t,\vc{x}\right) v^i\left(t,\vc{x}\right) v^j\left(t,\vc{x}\right) + \mathcal{P}\left(t,\vc{x}\right)\delta_{ij} + \mathcal{O}\left(c^{-2}\right). \nonumber
\end{eqnarray}
We express the spatial integrals over the source in (\ref{lingmunusol}) and (\ref{linphisolexp}) using the following momenta of the energy--momentum tensor,
\begin{equation}\label{momenta1}
M^{I_n}(t) := \frac{1}{c^2}\int_{\mathbb{R}^3}d^3x\,T^{00}\left(t,\vc{x}\right)x^{I_n}, \hspace{1.3cm}
S^{ijI_n}(t) := \int_{\mathbb{R}^3}d^3x\,T^{ij}\left(t,\vc{x}\right)x^{I_n},
\end{equation}
and the quantities
\begin{eqnarray}\label{momenta2}
\mathcal{M}^{I_n}_{klm}(t) &:=& \int_\mathbb{R}ds\,q(s)\frac{p^k(s)}{r^lc^m}\frac{\partial^m}{\partial t^m}
M^{I_n}(\tau(t,s)), \\
\mathcal{S}^{ijI_n}_{klm}(t) &:=& \int_\mathbb{R}ds\,q(s)\frac{p^k(s)}{r^lc^m}\frac{\partial^m}{\partial t^m}
S^{ijI_n}(\tau(t,s)). \nonumber
\end{eqnarray}
Here $I_n$ denotes a string of indices $i_1\ldots i_n$, $n=0,1,2\ldots$, and $x^{I_n} := x^{i_1}\cdots x^{i_n}$. Moreover, we will denote $\mathcal{M} := \mathcal{M}_{000}$. It is useful to introduce 
the following linear combinations of the quantities (\ref{momenta2}),
\begin{eqnarray}\label{comb}
D^i(t) &:=& \mathcal{M}_{210}^i(t) + \mathcal{M}_{101}^i(t),\\
Q_1^{ij}(t) &:=& \frac{3}{2}\mathcal{M}_{420}^{ij}(t) + \mathcal{M}_{311}^{ij}(t) 
+ \frac{1}{2}\mathcal{M}_{202}^{ij}(t), \nonumber\\
Q_2^{ij}(t) &:=& {}-\mathcal{S}_{000}^{ij}(t) - \frac{1}{2}\mathcal{M}_{220}^{ij}(t) 
- \frac{1}{2}\mathcal{M}_{111}^{ij}(t),\nonumber\\
O_1^{ijk}(s) &:=& \frac{5}{2}\mathcal{M}_{630}^{ijk}(t) + \frac{1}{2}\mathcal{M}_{521}^{ijk}(t) + \frac{1}{2}\mathcal{M}_{412}^{ijk}(t) + \frac{1}{6}\mathcal{M}_{303}^{ijk}(t), \nonumber\\
O_2^{ijk}(s) &:=& {}-\mathcal{S}_{210}^{ijk}(t) - \mathcal{S}_{101}^{ijk}(t) 
- \frac{3}{2}\mathcal{M}_{430}^{ijk}(t) - \frac{1}{2}\mathcal{M}_{321}^{ijk}(t) 
- \mathcal{M}_{212}^{ijk}(t), \nonumber\\
H_1^{ijkl}(s) &:=& \frac{35}{8}\mathcal{M}_{840}^{ijkl}(t) + \frac{35}{8}\mathcal{M}_{731}^{ijkl}(t)
+ \frac{15}{8}\mathcal{M}_{622}^{ijkl}(t) - \frac{1}{12}\mathcal{M}_{513}^{ijkl}(t) 
+ \frac{1}{24}\mathcal{M}_{404}^{ijkl}(t), \nonumber\\
H_2^{ijkl}(s) &:=& {}-\mathcal{S}_{420}^{ijkl}(t) - \mathcal{S}_{311}^{ijkl}(t) - \mathcal{S}_{202}^{ijkl}(t) \nonumber\\
& & \quad {}- \frac{15}{4}\mathcal{M}_{640}^{ijkl}(t) - \frac{15}{4}\mathcal{M}_{531}^{ijkl}(t) 
- \frac{3}{2}\mathcal{M}_{422}^{ijkl}(t) + \frac{1}{12}\mathcal{M}_{313}^{ijkl}(t), \nonumber\\
H_3^{ijkl}(s) &:=& {}-\mathcal{S}_{220}^{ijkl}(t) - \mathcal{S}_{111}^{ijkl}(t) 
- \frac{3}{8}\mathcal{M}_{440}^{ijkl}(t) - \frac{3}{8}\mathcal{M}_{331}^{ijkl}(t) 
- \frac{1}{8}\mathcal{M}_{222}^{ijkl}(t).\nonumber
\end{eqnarray}
Taking into account up to quadrupoles, and hexadecapoles, for $\gamma^{ij}$, and $\varphi$, respectively, the asymptotic fields (\ref{lingmunusol}) and (\ref{linphisolexp}) can be written in terms of the quantities (\ref{momenta1}), (\ref{momenta2}) and (\ref{comb})
\begin{eqnarray}
\label{linsolmm}
\gamma^{ij}(t,\vc{x}) & = & 
\frac{2G}{c^4r}\frac{\partial^2}{\partial t^2}M^{ij}(t-r/c), \\
\label{linsolms} 
\varphi(t,\vc{x}) & = & -\frac{G}{3ac^2r}\Bigg[
\mathcal{M}(t) + n_i\,D^i(t) + n_in_j\,Q_1^{ij}(t) + \delta_{ij}\,Q_2^{ij}(t)  \\
& & \quad {}+ n_in_jn_k\,O_1^{ijk}(t) +
n_i\delta_{jk}\,O_2^{ijk}(t) \nonumber  \\
& & \quad {}+ n_in_jn_kn_l\,H_1^{ijkl}(t) + n_in_j\delta_{kl}\,H_2^{ijkl}(t) 
+ \delta_{ij}\delta_{kl}\,H_3^{ijkl}(t)\Bigg], \nonumber
\end{eqnarray}
where we have used the same symbols $\gamma^{ij}$ and $\varphi$ for the approximations.
From (\ref{momenta2}) we infer that the dimensionless field $a\varphi$ depends on $a$ 
through the function $p$. This irrational dependence on $r$ arises by the same reason 
as the Yukawa like terms $e^{-r/\sqrt{6a} }$ in the solutions for $\varphi$ for an isolated 
system, since the field $\varphi$ has a range $1/\sqrt{6a}$ as per equation (\ref{eullag2}).
In the same way as in the $1/c$ expansion of the metric field \cite{phich,clifton}, 
the approximation is parametrised by a dimensionless parameter $a/\ell^2$, where $\ell$ is a 
typical length depending on the model. As for the $1/c$ expansion of the field of a 
quasi static isolated source, also in the present 
case $\ell$ corresponds to the distance from the source $r$, i.e.~the distance the wave has 
propagated through. Consequently, even the field $ra\varphi$ converges to zero for 
$r \to \infty$, that is, other than the pure metric field (\ref{linsolmm}), 
the scalar field decays faster than $1/r$. This is expected since the scalar field 
is massive, leading to a nontrivial dispersion relation for a propagating wave. 


\section{The Energy--Momentum Complex}

In order to derive the energy flux of a gravitational wave in $f(R)$ inspired scalar 
tensor theory, we need an analogue to the Landau--Lifshitz complex $t^{\mu\nu}_{LL}$ in 
GR \cite{LL}. This can be obtained by using the method in \cite{NUTKU}, where an energy-momentum 
complex in the Brans--Dicke theory is presented. This method applies to theories 
derived from convex or concave functions $f$, i.e.~theories that are equivalent to scalar--tensor 
theories. In this section, we consider a general metric tensor $g_{\mu\nu}$ and scalar field $\phi$. 
In particular, we do not constrain the asymptotic values of these fields. 
For an alternative derivation of the energy--momentum complex
cf.~\cite{begair,steyun}.

Defining 
\begin{equation}\label{lefths}
X^{\mu\nu} := R^{\mu\nu} - \frac{1}{2}R g^{\mu\nu} + \frac{1}{8a\phi}(\phi-1)^2g^{\mu\nu}
-\frac{1}{\phi}\left(\nabla^{\mu}\nabla^{\mu}\phi - g^{\mu\nu}\square_g \phi\right)
\end{equation}
and
\begin{equation}\label{umnls}
U^{\mu\nu\lambda\sigma} := \phi^2(-g)
\left(g^{\mu\nu}g^{\lambda\sigma} - g^{\mu\sigma}g^{\lambda\nu}\right),
\end{equation}
we can write the generalisation of the Landau--Lifshitz complex as
\begin{equation}\label{emcomp}
t^{\mu\nu} = \frac{c^4\phi}{8\pi G}
\left[\frac{1}{2\phi^2(-g)}\partial_\lambda \partial_\sigma U^{\mu\lambda\nu\sigma} 
- X^{\mu\nu}\right].
\end{equation}
By construction, the energy--momentum conservation laws then can be cast into the form
\begin{equation}
\partial_\mu\left[\phi(-g)\left(T^{\mu\nu} + t^{\mu\nu}\right)\right] = 0.
\end{equation}
Using (\ref{lefths}) and (\ref{umnls}), it is straight forward to express 
the energy momentum complex (\ref{emcomp}) in terms of the fields 
$g_{\mu\nu}$, $\phi$, their first partial derivatives and the connection coefficients 
$\Gamma^\lambda_{\mu\nu}$ as
\begin{eqnarray}\label{emcompf}
t^{\mu\nu} & = & \phi t^{\mu\nu}_{LL} + \frac{c^4}{16\pi G\phi}
\Bigg[2g^{\mu\nu}\partial_\lambda\phi\partial^\lambda\phi - 
2\partial^\mu\phi\partial^\nu\phi - \frac{1}{4a}(\phi-1)^2g^{\mu\nu}\Bigg] \\
& & \quad {}+ \frac{c^4}{8\pi G}\Bigg[
g^{\mu\nu}\left(2\partial^\lambda\phi\Gamma^\sigma_{\lambda\sigma} 
- \partial_\lambda\phi\Gamma^\lambda_{\sigma\rho}g^{\sigma\rho}\right) + 
g^{\mu\lambda}g^{\nu\sigma}\partial_\rho\phi\Gamma^\rho_{\lambda\sigma} \nonumber \\
& & \quad {}+ \left(\partial^\mu\phi\Gamma^\nu_{\lambda\sigma} 
+ \partial^\nu\phi\Gamma^\mu_{\lambda\sigma}\right)g^{\lambda\sigma}
- \left(\partial^\mu\phi g^{\nu\lambda} 
+ \partial^\nu\phi g^{\mu\lambda}\right)\Gamma^\sigma_{\lambda\sigma} \nonumber \\
& & \quad {}- \left( g^{\mu\lambda} 
\Gamma^\nu_{\lambda\sigma}
+ g^{\nu\lambda}\Gamma^\mu_{\lambda\sigma}\right)\partial^\sigma\phi
\Bigg]. \nonumber
\end{eqnarray}
The energy flux in an arbitrary direction $x^i$ is given by the component $t^{0i}$.


\section{Energy Emission of Isolated Systems}

Consider a plane gravitational wave propagating in vacuum in the $x^i$ direction. 
In addition to the gauge (\ref{gauge}), it is possible to perform a further gauge 
transformation which makes the $\varphi$ independent part of the perturbation transverse 
and traceless (TT) \cite{caplau,CORDA1,begair}, such that we can write   
\begin{equation}
h^{\mu\nu}(t-x^i/c) = \gamma^{\mu\nu}_{TT_i}(t-x^i/c) - 2a\eta^{\mu\nu}\varphi(t,x^i)
\end{equation}
In this gauge, we evaluate the energy flux to leading order 
in the perturbation fields $\gamma_{\mu\nu}$ and $\varphi$. By angle brackets we denote 
the average over a four-dimensional spacetime region with an extension which is much larger than a 
typical wavelength. The formula (\ref{emcompf}) then yields
\begin{equation}\label{flux}
t^{0i} = \frac{c^4}{32\pi G}\left<\partial_0\gamma^{jk}_{TT_i}
\partial_0\gamma^{jk}_{TT_i} + 24a^2(\partial_0\varphi)^2\right>.
\end{equation}
The first term on the right hand side of (\ref{flux}) can be evaluated in the same way as in 
GR by means of (\ref{linsolmm}) and the trace--free quadrupole tensor 
\begin{equation}
\mathcal{Q}^{ij}(t) = \frac{1}{c^2}\left(3M^{ij} - \delta_{ij}M^{kk}\right),
\end{equation}
From (\ref{linsolms}) 
we obtain for the second term on the right hand side of (\ref{flux}) the asymptotic value
\begin{eqnarray}
24a^2(\partial_0\varphi)^2 & = & \frac{8G^2}{3c^6r^2}
\Bigg[\dot{\mathcal{M}}^2
+ 2n_i\dot{\mathcal{M}}\dot{D}^i + n_in_j\left(2\dot{\mathcal{M}}\dot{Q}_1^{ij} 
+ \dot{D}^i\dot{D}^j\right) 
+ 2\delta_{ij}\dot{\mathcal{M}}\dot{Q}_2^{ij}  \\
& & \quad {}+ 2n_in_jn_k\left(\dot{\mathcal{M}}\dot{O}_1^{ijk} + 
\dot{D}^i\dot{Q}_1^{jk}\right) - 2n_i\delta_{jk}\left(\dot{\mathcal{M}}\dot{O}_2^{ijk} 
+ \dot{D}^i\dot{Q}_2^{jk}\right) \nonumber \\
& & \quad {}+ 
n_in_jn_kn_l\left(2\dot{\mathcal{M}}\dot{H}_1^{ijkl} 
+ 2\dot{D}^{i}\dot{O}_1^{jkl} 
+ \dot{Q}_1^{ij}\dot{Q}_1^{kl}\right)Ê\nonumber\\
& & \quad {}+ 2n_in_j\delta_{kl}\left(\dot{\mathcal{M}}\dot{H}_2^{ijkl} 
+ \dot{D}^i\dot{O}_2^{jkl} + \dot{Q}_1^{ij}\dot{Q}_2^{kl}\right)\Big) \nonumber\\
& & \quad {}+ \delta_{ij}\delta_{kl}\left(2\dot{\mathcal{M}}\dot{H}_3^{ijkl} 
+ \dot{Q}_2^{ij}\dot{Q}_2^{kl}\right)\Big)\Bigg]. \nonumber
\end{eqnarray}
The total power 
of the source is obtained by integrating (\ref{flux}) over a sphere with radius $r$:
\begin{eqnarray}\label{multipolform}
P & = & \frac{G}{3c}\Bigg<\dot{\mathcal{M}}^2
+  \frac{1}{3}\left[\dot{\mathcal{M}}\left(2\dot{Q}_1^{ij} + 6\dot{Q}_2^{ij}\right) 
+ \dot{D}^i\dot{D}^j\right] \\ 
& & \hspace{1cm} {}+ \frac{1}{15}\Big[\dddot{\mathcal{Q}}^{ij}\dddot{\mathcal{Q}}^{ij}
+ \dot{\mathcal{M}}\left(6\dot{H}_1^{iijj} + 10\dot{H}_2^{iijj} + 30\dot{H}_3^{iijj}\right) 
+ \dot{D}^{i}\left(6\dot{O}_1^{ijj} + 10\dot{O}_2^{ijj}\right)\nonumber\\
& & \hspace{2.5cm} {}+ \dot{Q}_1^{ii}\left(\dot{Q}_1^{jj} - 10\dot{Q}_2^{jj}\right) + 2\dot{Q}_1^{ij}\dot{Q}_1^{ij} + 15\dot{Q}_2^{ij}\dot{Q}_2^{ij}
\Big] \nonumber
\Bigg>,
\end{eqnarray}
where the dot denotes the derivative with respect to $t$.
Note that the additional degree of freedom represented by the scalar field predicts 
radiation of all multipoles, in particular monopoles and dipoles. These are of 
lower order than the original quadrupole contribution in GR represented by the moments 
$\mathcal{Q}^{ij}$, and could principally lead 
to a non negligible contribution in concrete applications. Since the quantities 
$\mathcal{M}^I$ and $\mathcal{S}^I$ as well as their derivatives vanish in the limit 
$r \to \infty$, the energy radiated by means of the scalar degree of freedom 
is absorbed as the field $\varphi$ disperses completely when propagating to timelike 
infinity.


\section{Applications}

The most interesting application of the formal result (\ref{multipolform}) 
is the energy loss of binary systems by the emission of gravitational radiation, 
in particular binary pulsars such as 
the PSR J0737-3039 system \cite{Lyn,Bur}. In quadratic $f(R)$ gravity, 
the non--relativistic motion of compact objects is governed by the Newtonian 
potential with an additional Yukawa correction. This implies 
that the Keplerian orbits also need 
appropriate corrections, cf.~for example \cite{phich}. A general treatment 
including generic orbits is beyond the scope of this work. In order to obtain 
a rough estimate for the correction terms in (\ref{multipolform}), we apply 
the formula to the radiation of binaries in circular orbits. 
For this application we expect that the total mass of the system changes on a 
time scale which is much larger than the orbital period. Hence, the monopole 
contribution to (\ref{multipolform}) is negligible, $\dot{\mathcal{M}} \approx 0$, 
whereas for appropriate mass densities the dipole 
term is the leading order contribution. 

Consider a binary system consisting of two point masses $m_1$ and $m_2$ 
in a circular orbit moving at angular velocity $\omega$. For $m_1 \ne m_2$, 
the dipole moment does not vanish. The only non--relativistic correction 
is a modification of Kepler's third law \cite{phich},
\begin{equation}\label{kepler}
\omega^2 = \frac{G(m_1+m_2)}{d^3}\left[1 + 
\frac{1}{3}\left(1 + \frac{d}{\sqrt{6a}}\right)e^{-d/\sqrt{6a}}\right],
\end{equation}
which can be numerically solved for the orbital separation $d$. This 
non--relativistic correction is parametrised by $a/d^2$, 
while the corrections arising from the propagation of the field $\varphi$ 
are a function of $a/r^2$. The choice of the ratio $d/r$ is constrained to 
be small for the multipole expansion to be viable.

We choose coordinates such that the motion is restricted to the $(x^1,x^2)$--plane. 
In a corotating frame, the mass density can be written as
\begin{equation}
\rho(\vc{x}) = \delta(x^2)\delta(x^3)\left[m_1\delta\left(x^2-d/2\right) + m_2\delta\left(x^2+d/2\right)\right].
\end{equation}
The dipole moments are therefore given by
\begin{equation}
M^1(t) = \frac{d(m_1 - m_2)\cos(\omega t)}{2}, \quad M^2(t) = M^1\left(t-\frac{\pi}{2\omega}\right),
\quad M^3(t) = 0,
\end{equation}
leading to
\begin{eqnarray}
\mathcal{M}^1_{klm}(t) &=& \frac{d(m_1 - m_2)}{2}\int_\mathbb{R}ds\,\frac{p^k(s)}{r^lc^m}\,q(s)\frac{\partial^m}{\partial t^m}\cos(\omega\tau(t,s)), \\
\mathcal{M}^2_{klm}(t) &=& \mathcal{M}^1_{klm}\left(t-\frac{\pi}{2\omega}\right). \nonumber
\end{eqnarray}
Taking the average of the dipole contribution over one period $\mathcal{T} := 2\pi/\omega$ we obtain
\begin{eqnarray}
P_d & = & \frac{G}{9c}\left<\left(\dot{D}^1\right)^2 + \left(\dot{D}^2\right)^2\right> \\
& = & \frac{G\omega}{9\pi c}\int_0^\mathcal{T} dt\left[\left(\dot{\mathcal{M}}^1_{210}(t)\right)^2 
+ \left(\dot{\mathcal{M}}^1_{101}(t)\right)^2 
+ 2\dot{\mathcal{M}}^1_{210}(t)\dot{\mathcal{M}}^1_{101}(t)\right]. \nonumber
\end{eqnarray}
\begin{figure}
\psfrag{a}{\footnotesize{$a\left[\mathrm{m}^2\right]$}}
\psfrag{P}{\footnotesize{$P_d \; [\mathrm{W}]$}}
\includegraphics[width=130mm]{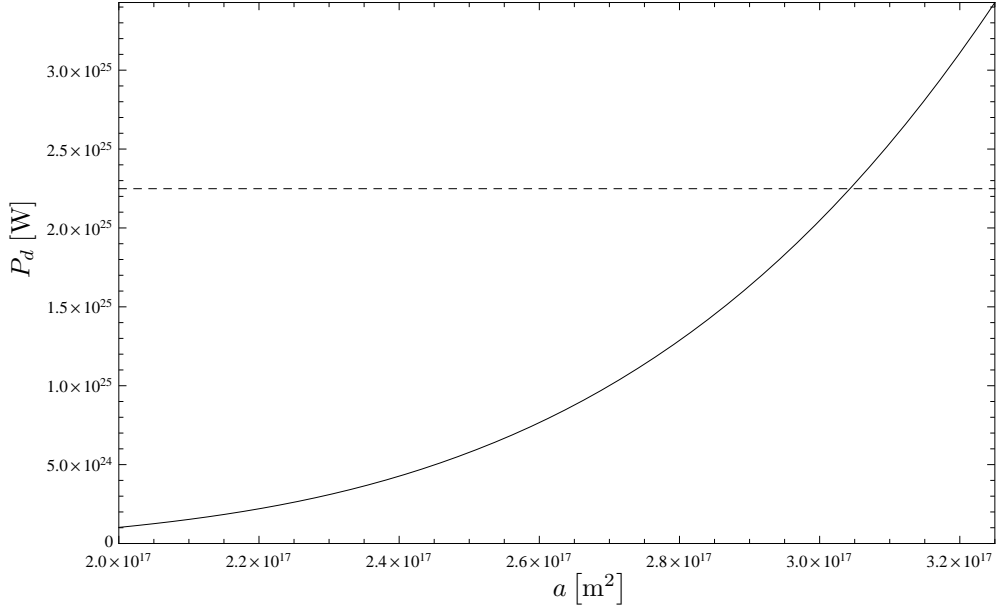}
\caption{Dipole power $P_d$ as a function of the model parameter $a$ for the data (\ref{data}). 
The value of the quadrupole power as predicted by GR, $P_{q}^{\textrm{GR}} = 2.248\cdot10^{25}\,\mathrm{W}$, 
is indicated by the dashed line.}
\end{figure}
\noindent In Figure 1 we plot $P_d$ against the model parameter $a$ for the data
\begin{align}\label{data}
m_1 &= 1.337 \, M_{\bigodot}, & m_2 &= 1.250 \, M_{\bigodot}, \\
\omega &= 0.7112\cdot10^{-4} \, \mathrm{s}^{-1}, & r &= 10^{10} \mathrm{m}, \nonumber
\end{align}
\noindent which is inspired by the PSR J0737-3039 system, but moving in a circular orbit 
with angular velocity $2\pi/\mathcal{T}$, where $\mathcal{T} = 8.834\cdot10^3 \, \mathrm{s}$ is the measured 
orbital period \cite{Lyn}. The parameter $a$ lies within the interval 
$\mathcal{I} = \left[2\cdot10^{17}\,\mathrm{m}^2,\,3.25\cdot10^{17}\,\mathrm{m}^2\right]$, such that 
the dipole term is comparable to the value of the quadrupole power as predicted by GR, 
$P_{q}^{\textrm{GR}} = 2.248\cdot10^{25}\,\mathrm{W}$. From (\ref{kepler}) 
we infer that $d \approx 9.5\cdot10^8 \, \mathrm{m}$ for $a \in \mathcal{I}$; the 
Newtonian value is $d = 8.798\cdot10^8 \, \mathrm{m}$. We choose the distance $r$ from 
the source such that $d/r \approx 1/10$.
For this data, the dipole power equals $1\%$ of the GR value at 
$a \approx 1.7\cdot10^{17}\,\mathrm{m}^2$.


\section{Conclusion}

We have derived the $f(R)$ correction terms to the GR quadrupole formula for the 
emission of gravitational radiation to leading order. An 
important result is that, in contrast to GR, quadratic $f(R)$ theory predicts the 
radiation of monopoles and dipoles. This is the case for nearly every alternative metric 
gravity theory known today \cite{WILL}. 

When considering a hypothetical binary similar to the double pulsar PSR J0737-3039, 
we found for the constant $a$ an upper limit of about $1.7\cdot10^{17}\,\mathrm{m}^2$. 
It would be interesting to study the result for real data of pulsars to see 
whether this limit can be improved. Notice that the limit on $a$ we got from the
geodetic precession using the Gravity Probe B data and the precession of the pulsar
PSR J0737-3039B is somewhat more stringent \cite{phich}.
When completing our paper we came aware of the recent publication 
\cite{begair} as well as the preprint \cite{CAP35}, which contain
equivalent results to some extent. The multipole expansion we focus on in our paper is, 
however, complementary to \cite{begair}, and the treatment in \cite{CAP35} is based 
on different assumptions concerning the conservation laws.

\begin{center}
{\bf Acknowledgements}
\end{center}

The authors wish to thank Norbert Straumann for useful discussions and the referee 
for some clarifying suggestions.


\begin{thebibliography}{20}
\expandafter\ifx\csname natexlab\endcsname\relax\def\natexlab#1{#1}\fi
\expandafter\ifx\csname bibnamefont\endcsname\relax
  \def\bibnamefont#1{#1}\fi
\expandafter\ifx\csname bibfnamefont\endcsname\relax
  \def\bibfnamefont#1{#1}\fi
\expandafter\ifx\csname citenamefont\endcsname\relax
  \def\citenamefont#1{#1}\fi
\expandafter\ifx\csname url\endcsname\relax
  \def\url#1{\texttt{#1}}\fi
\expandafter\ifx\csname urlprefix\endcsname\relax\def\urlprefix{URL }\fi
\providecommand{\bibinfo}[2]{#2}
\providecommand{\eprint}[2][]{\url{#2}}

\bibitem[{\citenamefont{{Will}}(1993)}]{WILL}
\bibinfo{author}{\bibfnamefont{C.~M.} \bibnamefont{{Will}}},
  \emph{\bibinfo{title}{{Theory and Experiment in Gravitational Physics}}}
  (\bibinfo{year}{1993}).

\bibitem[{\citenamefont{{Capozziello} et~al.}(2010)\citenamefont{{Capozziello},
  {de Laurentis}, and {Faraoni}}}]{caplau}
\bibinfo{author}{\bibfnamefont{S.}~\bibnamefont{{Capozziello}}},
  \bibinfo{author}{\bibfnamefont{M.}~\bibnamefont{{de Laurentis}}},
  \bibnamefont{and}
  \bibinfo{author}{\bibfnamefont{V.}~\bibnamefont{{Faraoni}}},
  \bibinfo{journal}{The Open Astronomy Journal} \textbf{\bibinfo{volume}{3}},
  \bibinfo{pages}{49} (\bibinfo{year}{2010}), \eprint{0909.4672}.

\bibitem[{\citenamefont{{Straumann}}(2011)}]{STR2}
\bibinfo{author}{\bibfnamefont{N.}~\bibnamefont{{Straumann}}},
  \bibinfo{journal}{To appear in a forthcoming volume in the series of Einstein
  Studies}  (\bibinfo{year}{2011}), \eprint{0809.5148}.

\bibitem[{\citenamefont{{Starobinsky}}(1980)}]{staro}
\bibinfo{author}{\bibfnamefont{A.~A.} \bibnamefont{{Starobinsky}}},
  \bibinfo{journal}{Physics Letters B} \textbf{\bibinfo{volume}{91}},
  \bibinfo{pages}{99} (\bibinfo{year}{1980}).

\bibitem[{\citenamefont{Vilenkin}(1985)}]{vilen}
\bibinfo{author}{\bibfnamefont{A.}~\bibnamefont{Vilenkin}},
  \bibinfo{journal}{Phys. Rev. D} \textbf{\bibinfo{volume}{32}},
  \bibinfo{pages}{2511} (\bibinfo{year}{1985}).

\bibitem[{\citenamefont{{Corda}}(2010)}]{CORDA1}
\bibinfo{author}{\bibfnamefont{C.}~\bibnamefont{{Corda}}},
  \bibinfo{journal}{European Physical Journal C} \textbf{\bibinfo{volume}{65}},
  \bibinfo{pages}{257} (\bibinfo{year}{2010}), \eprint{1007.4077}.

\bibitem[{\citenamefont{{Corda}}(2008{\natexlab{a}})}]{CORDA2}
\bibinfo{author}{\bibfnamefont{C.}~\bibnamefont{{Corda}}},
  \bibinfo{journal}{General Relativity and Gravitation}
  \textbf{\bibinfo{volume}{40}}, \bibinfo{pages}{2201}
  (\bibinfo{year}{2008}{\natexlab{a}}), \eprint{0802.2523}.

\bibitem[{\citenamefont{{Corda}}(2008{\natexlab{b}})}]{CORDA3}
\bibinfo{author}{\bibfnamefont{C.}~\bibnamefont{{Corda}}},
  \bibinfo{journal}{International Journal of Modern Physics A}
  \textbf{\bibinfo{volume}{23}}, \bibinfo{pages}{1521}
  (\bibinfo{year}{2008}{\natexlab{b}}), \eprint{0711.4917}.

\bibitem[{\citenamefont{{Berry} and {Gair}}(2011)}]{begair}
\bibinfo{author}{\bibfnamefont{C.~P.~L.} \bibnamefont{{Berry}}}
  \bibnamefont{and} \bibinfo{author}{\bibfnamefont{J.~R.}
  \bibnamefont{{Gair}}}, \bibinfo{journal}{\prd} \textbf{\bibinfo{volume}{83}},
  \bibinfo{pages}{104022} (\bibinfo{year}{2011}), \eprint{1104.0819}.

\bibitem[{\citenamefont{{Wagoner}}(1970)}]{wagon}
\bibinfo{author}{\bibfnamefont{R.~V.} \bibnamefont{{Wagoner}}},
  \bibinfo{journal}{\prd} \textbf{\bibinfo{volume}{1}}, \bibinfo{pages}{3209}
  (\bibinfo{year}{1970}).

\bibitem[{\citenamefont{{Khoury} and {Weltman}}(2004)}]{KhWe}
\bibinfo{author}{\bibfnamefont{J.}~\bibnamefont{{Khoury}}} \bibnamefont{and}
  \bibinfo{author}{\bibfnamefont{A.}~\bibnamefont{{Weltman}}},
  \bibinfo{journal}{Physical Review Letters} \textbf{\bibinfo{volume}{93}},
  \bibinfo{pages}{171104} (\bibinfo{year}{2004}),
  \eprint{arXiv:astro-ph/0309300}.

\bibitem[{\citenamefont{{N{\"a}f} and {Jetzer}}(2010)}]{phich}
\bibinfo{author}{\bibfnamefont{J.}~\bibnamefont{{N{\"a}f}}} \bibnamefont{and}
  \bibinfo{author}{\bibfnamefont{P.}~\bibnamefont{{Jetzer}}},
  \bibinfo{journal}{\prd} \textbf{\bibinfo{volume}{81}},
  \bibinfo{pages}{104003} (\bibinfo{year}{2010}), \eprint{1004.2014}.

\bibitem[{\citenamefont{{Clifton}}(2008)}]{clifton}
\bibinfo{author}{\bibfnamefont{T.}~\bibnamefont{{Clifton}}},
  \bibinfo{journal}{\prd} \textbf{\bibinfo{volume}{77}},
  \bibinfo{pages}{024041} (\bibinfo{year}{2008}), \eprint{0801.0983}.

\bibitem[{\citenamefont{{Morse} and {Feshbach}}(1953)}]{morfesh}
\bibinfo{author}{\bibfnamefont{P.~M.} \bibnamefont{{Morse}}} \bibnamefont{and}
  \bibinfo{author}{\bibfnamefont{H.}~\bibnamefont{{Feshbach}}},
  \emph{\bibinfo{title}{{Methods of Theoretical Physics}}}
  (\bibinfo{year}{1953}).

\bibitem[{\citenamefont{{Landau} and {Lifshitz}}(1992)}]{LL}
\bibinfo{author}{\bibfnamefont{L.~D.} \bibnamefont{{Landau}}} \bibnamefont{and}
  \bibinfo{author}{\bibfnamefont{E.~M.} \bibnamefont{{Lifshitz}}},
  \emph{\bibinfo{title}{{Klassische Feldtheorie}}}
  (\bibinfo{publisher}{Lehrbuch der theoretischen Physik.~Berlin:
  Akademie-Verlag, 12.~Auflage}, \bibinfo{year}{1992}).

\bibitem[{\citenamefont{{Nutku}}(1969)}]{NUTKU}
\bibinfo{author}{\bibfnamefont{Y.}~\bibnamefont{{Nutku}}},
  \bibinfo{journal}{\apj} \textbf{\bibinfo{volume}{158}}, \bibinfo{pages}{991}
  (\bibinfo{year}{1969}).

\bibitem[{\citenamefont{{Stein} and {Yunes}}(2011)}]{steyun}
\bibinfo{author}{\bibfnamefont{L.~C.} \bibnamefont{{Stein}}} \bibnamefont{and}
  \bibinfo{author}{\bibfnamefont{N.}~\bibnamefont{{Yunes}}},
  \bibinfo{journal}{\prd} \textbf{\bibinfo{volume}{83}},
  \bibinfo{pages}{064038} (\bibinfo{year}{2011}), \eprint{1012.3144}.

\bibitem[{\citenamefont{{Lyne}}(2006)}]{Lyn}
\bibinfo{author}{\bibfnamefont{A.~G.} \bibnamefont{{Lyne}}},
  \bibinfo{journal}{Chinese Journal of Astronomy and Astrophysics Supplement}
  \textbf{\bibinfo{volume}{6}}, \bibinfo{pages}{162} (\bibinfo{year}{2006}).

\bibitem[{\citenamefont{{Burgay} et~al.}(2006)\citenamefont{{Burgay},
  {D'Amico}, {Possenti}, {Manchester}, {Lyne}, {Kramer}, {McLaughlin},
  {Lorimer}, {Camilo}, {Stairs} et~al.}}]{Bur}
\bibinfo{author}{\bibfnamefont{M.}~\bibnamefont{{Burgay}}},
  \bibinfo{author}{\bibfnamefont{N.}~\bibnamefont{{D'Amico}}},
  \bibinfo{author}{\bibfnamefont{A.}~\bibnamefont{{Possenti}}},
  \bibinfo{author}{\bibfnamefont{R.~N.} \bibnamefont{{Manchester}}},
  \bibinfo{author}{\bibfnamefont{A.~G.} \bibnamefont{{Lyne}}},
  \bibinfo{author}{\bibfnamefont{M.}~\bibnamefont{{Kramer}}},
  \bibinfo{author}{\bibfnamefont{M.~A.} \bibnamefont{{McLaughlin}}},
  \bibinfo{author}{\bibfnamefont{D.~R.} \bibnamefont{{Lorimer}}},
  \bibinfo{author}{\bibfnamefont{F.}~\bibnamefont{{Camilo}}},
  \bibinfo{author}{\bibfnamefont{I.~H.} \bibnamefont{{Stairs}}},
  \bibnamefont{et~al.}, \bibinfo{journal}{Memorie della Societa Astronomica
  Italiana Supplementi} \textbf{\bibinfo{volume}{9}}, \bibinfo{pages}{345}
  (\bibinfo{year}{2006}).

\bibitem[{\citenamefont{{De Laurentis} and {Capozziello}}(2011)}]{CAP35}
\bibinfo{author}{\bibfnamefont{M.}~\bibnamefont{{De Laurentis}}}
  \bibnamefont{and}
  \bibinfo{author}{\bibfnamefont{S.}~\bibnamefont{{Capozziello}}},
  \bibinfo{journal}{ArXiv e-prints}  (\bibinfo{year}{2011}),
  \eprint{1104.1942}.

\end{thebibliography}
\end{document}